# Observation of cooperative strong coupling between optical phonon and crystal-field excitations in a pseudo Jahn-Teller system

**Authors:** Fangliang Wu[1, 2†], Xiaoxuan Ma[3, 4†], Zhongwei Zhang[1,2], Motoaki Bamba[5, 6],

Haowen Wu[3], Jian Sun[1,2], Yuan Wan[7, 8], Shixun Cao[3, 9*], Qi Zhang[1, 2*]

[1]National Laboratory of Solid State Microstructures and Department of Physics, Nanjing University, Nanjing 210093, China
[2]Jiangsu Physical Science Research Center, Nanjing University, Nanjing 210093, China
[3]Materials Genome Institute and Institute for Quantum Science and Technology, Shanghai University, Shanghai 200444, China
[4]Shanghai Key Laboratory of High Temperature Superconductors, Shanghai University, Shanghai, 200444, China
[5]Department of Physics, Graduate School of Engineering Science, Yokohama National University, Yokohama 240-8501, Japan
[6]Institute for Multidisciplinary Sciences, Yokohama National University, Yokohama 240-8501, Japan
[7]Institute of Physics, Chinese Academy of Sciences, Beijing 100190, China
[8]University of Chinese Academy of Sciences, Beijing 100049, China
[9] School of Material Science and Engineering, Shanghai University, Shanghai 200444, China

[†]These authors contributed equally to this work.

[*]Correspondence to: zhangqi@nju.edu.cn and sxcao@shu.edu.cn

**Abstract: Cooperative interactions between localized electronic excitations and crystal lattice are central to the emergence of complex structural phases in materials. However, the scaling relations governing these collective behaviors remain largely unexplored. Here, using magneto-Raman spectroscopy, we report the direct observation of the strongly coupled optical phonon and non-degenerate crystal-field excitations (CFEs) in $ErFeO_3$. By independently tuning the effective population of Jahn-Teller-active erbium ions through temperature and chemical dilution with Jahn-Teller-inactive yttrium ions, we identify the coupling strength varies linearly with the square root of electronic excitations population. Notably, Y-doping reveals the hybridization gap reduces significantly faster than predicted by density scaling alone, indicating phonon coherence is essential for establishing this cooperative interaction. Our findings highlight the role of optical phonons in mediating short-range interactions that drive cooperative Jahn-Teller effect, evidencing the pathway for tailoring electronic and vibrational properties of Jahn-Teller materials through population control.**

## MAIN TEXT

The Jahn-Teller effect (JTE) establishes a fundamental link between structural deformation and orbital physics. When local JT-active ions engage in cooperative coupling, which drives long-range lattice distortion and orbital ordering, the macroscopic properties of the material are profoundly altered. Such collective behavior culminates in a phase transition formally known as the cooperative JTE[1–3]. These interactions manifest in diverse phenomena, including giant magneto and electrostriction[4–6], metamagnetism characterized by complex phase diagrams[7–9], and the formation of JT polarons in systems exhibiting colossal magnetoresistance or unconventional superconductivity[10–13].

Conventionally, the cooperativity in JT ion ensembles is attributed to long-range strain fields and virtual phonon exchange mediated primarily by acoustic modes[4]. In contrast, optical phonons are often treated as local distortion fluctuations due to their short-range nature, leading to an underestimation of their contribution to cooperative JTE. However, while optical phonons possess low group velocities and relatively shorter lifetimes compared to acoustic modes, they play a critical role in establishing structural phases that go beyond simple ferro-distortive types, such as antiferrodistortion[4], helicoidal distortion[14], and incommensurate distortion[15]. Elucidating the scaling relations of these optical-phonon-mediated interactions is therefore essential for understanding complex distortions, for instance, Davydov splitting of JT ions[16], distortion fluctuation above transition temperature ($T_c$)[17], and the overestimated $T_c$ of cooperative JTE[18].

Experimentally, the measurements of specific heat, elastic modulus, and ultrasounds usually used in studying cooperative JTE are hard to reach the realm of optical phonons. By contrast, the coherent coupling between phonon and orbital excitations provides an opportunity to study their cooperative behavior via spectroscopic methods. The temperature and magnetic dependent splitting of both phonons and orbital states have been observed via Raman scattering and infrared absorption[19–23]. In addition, the avoided-crossing of coupled acoustic phonon-orbital excitation has also been observed via neutron scattering[24]. Nevertheless, the avoided crossing between optical phonons and orbital excitations, enabling the tracking of its cooperative coupling, remains largely unexplored.

In this paper, we report the direct observation of the scaling relation of cooperative interaction in pseudo-JTE involving optical phonon coupling, exemplified by a rare-earth orthoferrite $ErFeO_3$, with pseudo means the coupling between non-degenerate orbital states[2]. By performing magneto-Raman spectroscopy, we uncover coupled spin, lattice, and orbital degrees of freedom, including magnon-crystal-field excitation (CFE) strong coupling and phonon-CFE strong coupling. The latter exhibits a prominent spectral avoided-crossing between an optical phonon and the CFE of $Er^{3+}$. Crucially, we demonstrate that the hybridization gap scales with the square root of the effective population of the involved orbitals. Furthermore, by introducing JT-inactive yttrium dopants, we find that a 5 % doping level reduces the hybridization gap significantly faster the predicted $\sqrt{N}$-type density scaling alone, indicating the cooperative coupling between non-local optical phonon and local CFEs is fundamentally governed by phonon coherence. Theoretically, an $E\otimes\beta$ JT model with crystal field splitting is taken into account, we show a scaling relation of coupling strength that is similar to that of the Dicke model in quantum optics, which has also been studied recently in the context of magnonic superradiance[25–28], indicating general cooperativity in spin-Boson problems.

$ErFeO_3$ belongs to the orthorhombic space group *Pbnm*, with a distorted perovskite-type structure, as shown in Fig. 1(a). The $Fe^{3+}$ ions form a canted antiferromagnetic order below 640 K, while the Néel temperature ($T_N$) of $Er^{3+}$ is 4.5 K[29]. In addition, spins of $Fe^{3+}$ ions will reorient in the ac-plane within an interval temperature range ($T_1 < T < T_2$), as illustrated in Fig. 1(b). A unit cell of $ErFeO_3$ contains four $Fe^{3+}$ ions, corresponding to four magnon branches. Two of them are the acoustic magnons[30], i.e., the quasi-ferromagnetic mode (qFM) and the quasi-antiferromagnetic mode (qAFM), respectively. These two modes are observed in Raman measurements (Fig. 1(c)). As the temperature increases, the qFM mode exhibits a redshift accompanied by an increase in intensity until the spin reorientation regime, then it almost disappears at the high temperature $\Gamma_4$ phase. By contrast, the qAFM mode is only observed in the $\Gamma_4$ phase. Two frequency minima of the qFM mode have been observed, as shown in Fig. 1(d), corresponding to the spin reorientation temperatures $T_1$ (87 K) and $T_2$ (99 K), respectively. In addition, the qFM mode exhibits an intensity enhancement near $T_1$, which is due to the increased thermal population (Fig. 1(e))[31].

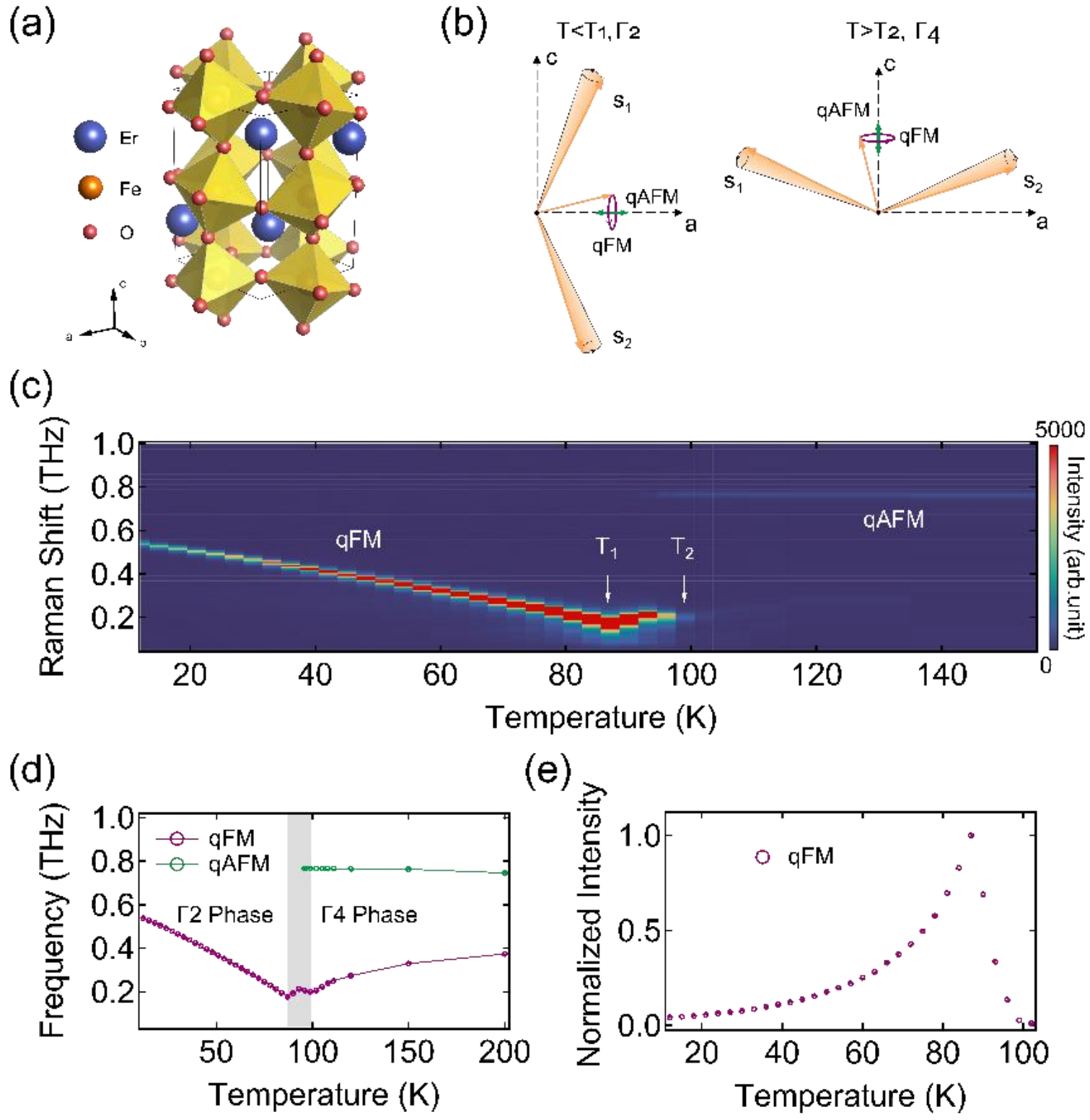

FIG. 1. Raman detection of acoustic magnons and spin reorientations in $ErFeO_3$. (a) The crystal structure of $ErFeO_3$. (b) Schematics of magnons and spin reorientation. $S_1$ and $S_2$ denote spins of $Fe^{3+}$ ions at neighboring

sublattices. The magenta and green arrows indicate the two types of collective precession of the net magnetic moment associated with qFM and qAFM modes, respectively. (c) Raman spectra of acoustic magnons as a function of temperature. Two frequency minima of the qFM mode are observed, and the corresponding temperatures are the lower ($T_1$) and upper ($T_2$) bounds of the spin reorientation range. (d), (e) The extracted peak position and intensity of qFM and qAFM as a function of temperatures. Error bars represent the standard error derived from the Lorentzian fitting.

The spin reorientation in $ErFeO_3$ originates from the temperature-dependent magnetic anisotropy[32], which relies on the orbital excitations of ground state multiplet of $Er^{3+}$. Due to the non-centrosymmetric Er sublattice (site symmetry $C_s$), all dipolar optical transitions among the $^4I_{15/2}$ ground state multiplet are allowed. Besides, it guarantees the symmetry requirement because the JT ions at inversion centers cannot couple to optical phonons at the $\boldsymbol{k} = 0$ point with odd-parity. From an experimental perspective, Raman spectroscopy detects $\boldsymbol{k} \approx 0$ modes due to the small momentum of photons and is sensitive to orbital and lattice excitations. The electronic Raman transition of rare-earth ions via electron-vibrational coupling has been theoretically proposed and experimentally verified[33–35]. Taken together, these factors provide the opportunity to observe the (pseudo-)JTE with optical phonon coupling.

In principle, the strong spin-orbit coupling (SOC) and weak crystal-field interaction of $Er^{3+}$ 4f electrons result in the hybridized Russell-Saunders states and the Stark splitting of the ground state $^4I_{15/2}$. Owing to the Er-Fe exchange interaction, the Kramers degeneracy is removed at zero magnetic field. It can exhibit further splitting under finite magnetic fields via Zeeman effect, as schematically shown in Fig. 2(d). By performing magneto-Raman spectroscopy on a c-cut $ErFeO_3$ crystal up to 9 T, multiple elementary excitations are observed below 4.5 THz, as presented in Fig. 2(a). The qFM and qAFM modes are marked by white arrows. Although absent at zero field, the qAFM mode gradually becomes detectable as the magnetic field increases. Besides magnons, the spin-flip CFE within Kramers doublets |±15/2> is observed with a g-factor of 5, which can be viewed as the electron paramagnetic resonance (EPR) of $Er^{3+}$ ions. Crystal-field excitations between |±15/2> and |±13/2> also appear around 1.35 THz. As shown in Fig. 2(d), there are four CFEs between two Stark energy levels, labeled as $E_{23}$, $E_{24}$, $E_{13}$, and $E_{14}$, with three of them being observed, matching previous reports[29, 36]. The g-factor of |±13/2> along the c-axis is 3.3. The effective field exerted by the Er-Fe exchange interaction on the |±13/2> states is also directly measured to be 0.13 THz at 10 K, which corresponds to the energy difference between $E_{13}$ and $E_{14}$ at zero field, as shown in Fig. 2(a). In addition, the broad peak around 2.04 THz is due to the transition from thermally populated |±13/2> to |±11/2> states.

The magnons and CFEs are strongly coupled in $ErFeO_3$, evidenced by the two avoided-crossing regimes. The first one is between the qFM and the $Er^{3+}$ EPR, accompanied by the enhancement of the Raman intensity of qFM at the high magnetic field branch. This observation is consistent with a previous THz absorption measurement[25]. The second one is between the qAFM and the CFE from |−15/2> to |−13/2>. The enhanced Raman spectral weight of qAFM can be attributed to hybridization with CFE. We present the peak positions of magnons and CFEs at different magnetic fields in Fig. 2(b). Phenomenologically, a 2 × 2 matrix can describe the coupling between magnon and CFE, with diagonal terms being uncoupled frequency and off-diagonals being coupling strength. At 10 K, the coupling strength is extracted to be 0.05 and 0.02 THz for qFM-EPR and qAFM-CFE couplings, respectively.

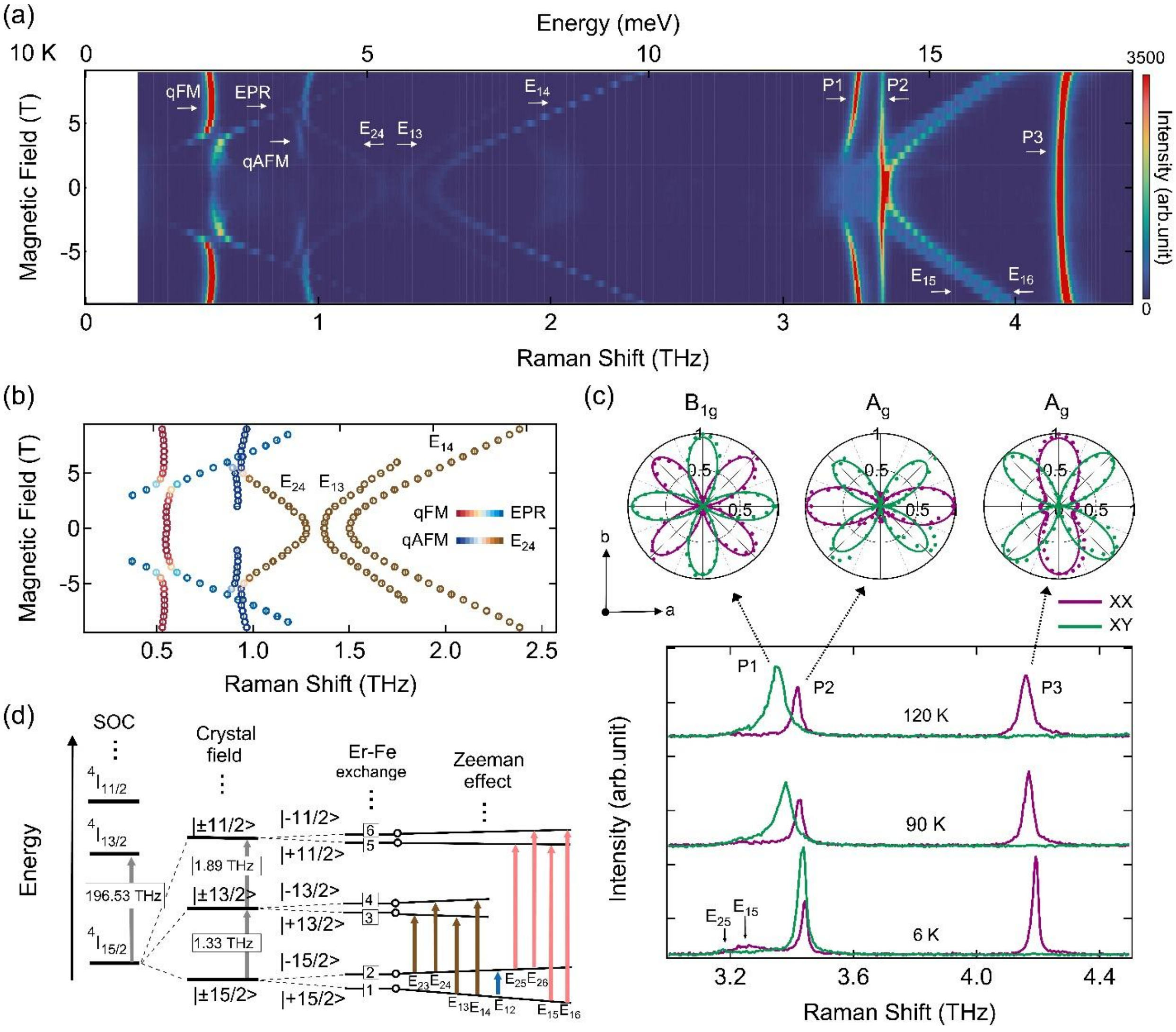

FIG. 2. Magneto-Raman spectra reveal coupled magnons, crystal-field excitations, and phonons in $ErFeO_3$. (a) At 10 K, magneto-Raman spectra of c-cut $ErFeO_3$ from −9 to +9 T. EPR represents electronic paramagnetic resonance. The $P_1$, $P_2$, and $P_3$ denote three phonons. The CFEs are labeled according to (d). (b) The extracted peak positions of magnons and CFEs. The colors of the data points indicate the proportion of different elementary excitations. (c) Polarization-resolved Raman spectra of $P_1$, $P_2$, and $P_3$ phonons. Vertical shift for clarification.

The magenta and green lines represent the Raman spectra in the co- (XX) and cross- (XY) polarization channels. (d) Schematic diagrams of energy levels of the $Er^{3+}$ ion.

Importantly, the sought-for optical phonon-CFEs strong coupling is observed around 3.4 THz. The modes slightly affected by the magnetic field are optical phonons of $ErFeO_3$[31], referred to as $P_1$ (3.3 THz), $P_2$ (3.4 THz), and $P_3$ (4.2 THz) hereinafter. Combining temperature-dependent and polarization-resolved Raman spectroscopy, we identify that the $P_1$, $P_2$, and $P_3$ phonons carry $B_{1g}$, $A_g$, and $A_g$ irreducible representations of point group $D_{2h}$, respectively, as shown in Fig. 2(c) (see details in supplementary text). The two modes with significant Zeeman shift correspond to the CFEs between the ground-state |+15/2> and the second excited state |±11/2>, labeled as $E_{15}$ and $E_{16}$, as shown in Fig. 2(d). The transition energy is consistent with previous absorption measurements[29]. The Zeeman splitting diagram of crystal-field transitions from |±15/2> to |±11/2> is presented in Fig. 3(a). The deviation from linearity at low magnetic fields results from the Er-Fe effective field. Besides a small g-factor along the c-axis, we find that |±11/2> also exhibits the quadratic Zeeman effect, namely, the overall quadratic blueshift with increasing magnetic field[25].

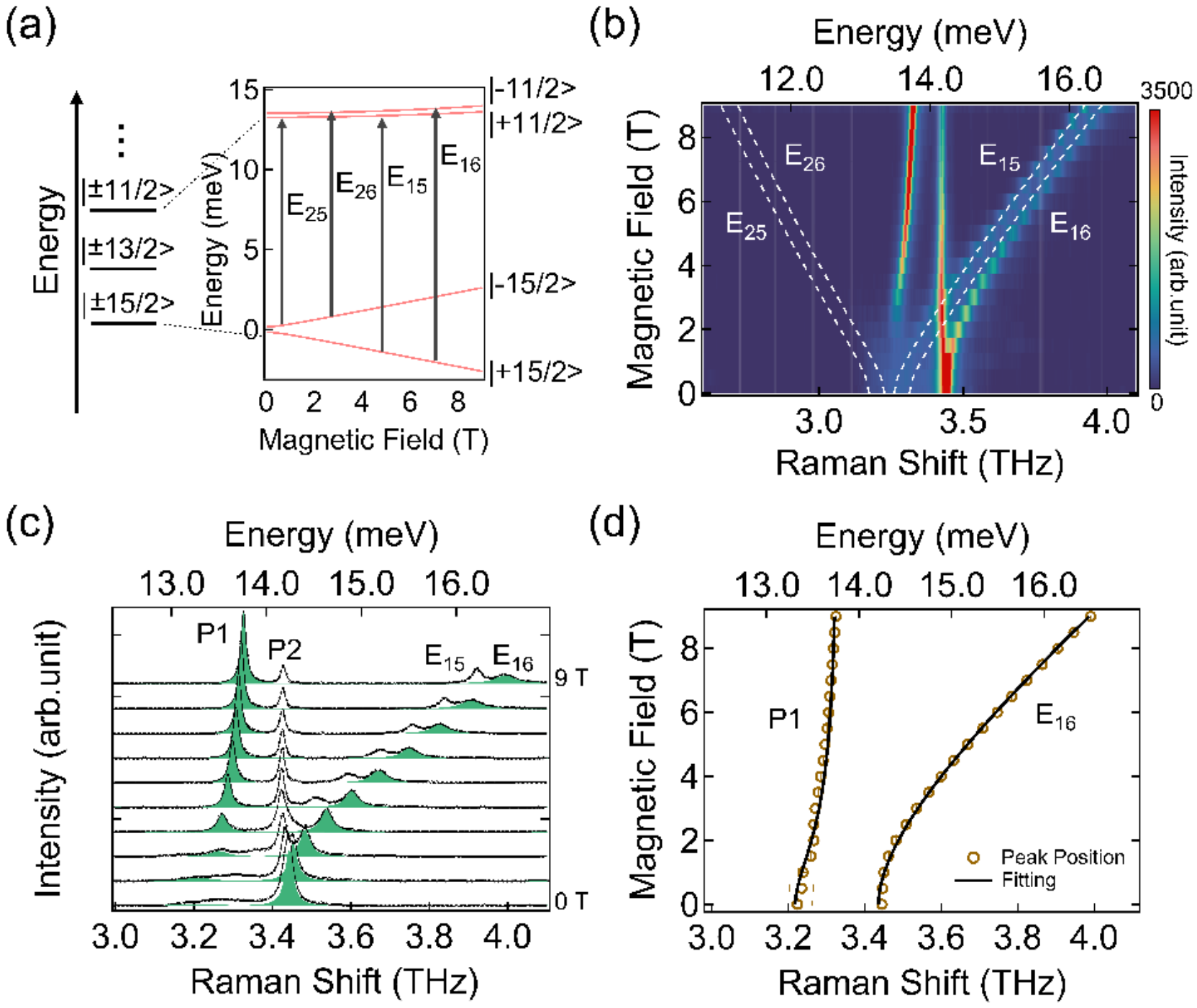

FIG. 3. Strongly coupled crystal-field excitation and P1 phonon at 10 K. (a) Schematic diagram of Zeeman-split crystal-field transitions from ground-state |±15/2> to the second excited state |±11/2>, labeled as $E_{25}$, $E_{26}$, $E_{15}$, and $E_{16}$. (b) Comparison between uncoupled CFEs (white dotted lines) and the experimental data. (c) Raman spectra of the phonon- CFE coupling region from 0 to 9 T. The green shaded peaks denote the modes that participate in the phonon-CFE coupling with Lorentzian fitting. (d) Extracted peak positions of $P_1$ and $E_{16}$ in (c). The fitting results of the 2 × 2 coupling matrix are shown by black lines.

The strong coupling between optical phonon and CFEs opens a hybridization gap, manifested as a spectral avoided-crossing. This feature is observed between $P_1$ and $E_{16}$ with spectral weight exchange at low magnetic fields. In contrast, $\mathrm{E}_{15}$ exhibits an inconspicuous avoided-crossing with $P_1$. Instead, at high magnetic fields, $E_{15}$ couples with $P_3$, as indicated by the increased spectral weight of $E_{15}$ and the slight blueshift of $P_3$ (see in Fig. 2(a)). From a symmetry perspective, the Kramers pair |+11/2> and |−11/2> carry different irreducible representations of the double point group $C_s$, meaning that $E_{15}$ and $E_{16}$ should behave oppositely under the mirror operation. This explains why only $P_1$-$E_{16}$ exhibits obvious coupling, while the $P_1$-$E_{15}$ coupling is symmetry-forbidden. Meanwhile, $P_2$ is slightly affected by $E_{15}$ and $E_{16}$. Similar to the analysis of magnon-CFE coupling, we use a 2×2 matrix to model the coupling between $P_1$ and $E_{16}$. The uncoupled frequencies of CFEs are depicted in Fig. 3(b) by white-dotted lines. Figure 3(c) presents the Lorentzian peak fitting of coupled phonon-CFE Raman spectra. The extracted eigenfrequencies are represented by black lines in Fig. 3(d), yielding a coupling strength of 0.11 THz at 10 K. A detailed description of the fitting procedure and symmetry analysis is provided in the supplementary information.

To investigate the cooperative nature of the phonon-CFE coupling, we utilize temperature as a control parameter for the effective population of local JTactive ions. As the thermal population increases, fewer JT ions participate in the cooperative coupling. Consequently, the hybridization gap gradually shrinks, causing the phonon and CFE frequencies to revert to their uncoupled values. As shown in Fig. 4(a), the eigenfrequencies derived from the coupling matrix (white dashed lines) show excellent agreement with the experimental Raman spectra. To quantify the scaling relation, we plot the extracted coupling strength, γ, as a function of $\sqrt{\eta_{CFE}}$. Here, $\sqrt{\eta_{CFE}}$ represents the effective ground state population, defined as the difference in occupation probabilities between |+15/2> and |−11/2>, which obeys the Boltzmann distribution at equilibrium (supplementary text). Remarkably, we observed a linear scaling relation between γ and $\sqrt{\eta_{CFE}}$ for the $P_1$-$E_{16}$ coupling, over a wide temperature range, as shown in Fig. 4(d) *left panel*. The results provide experimental evidence consistent with cooperative coupling among local $Er^{3+}$ ions mediated by an optical phonon field. This interaction follows a $\sqrt{N}$-type scaling relation, where $N$ represents the number of ions participating in the process.

Alternatively, chemical substitution offers another method to modulate $N$ by diluting the JT-active $Er^{3+}$ population with JT-inactive $Y^{3+}$ ions. As shown in Fig. 4(b), a 5 % Y-doping significantly broadens the phonon linewidth and induces a prominent reduction in the hybridization gap. While we observe a consistent linear scaling relation in $Er_{0.95}Y_{0.05}FeO_3$ (Fig. 4(d) *right panel*), the 5 % reduction in $N$ would theoretically predict only a 2.5 % decrease in the gap according to simple $\sqrt{N}$ scaling. This predicted value (dotted line) is only a quarter of the measured 10.4 % reduction, indicating that optical phonon coherence plays an essential role in JT-type electron-phonon coupling. Note that the $Y^{3+}$ doping brings both homogeneous and inhomogeneous broadenings of phonon linewidth, which underestimates the phonon intrinsic lifetime, but the significantly broadening of phonon linewidth gives qualitatively consistent to the large reduction of hybridization gap.

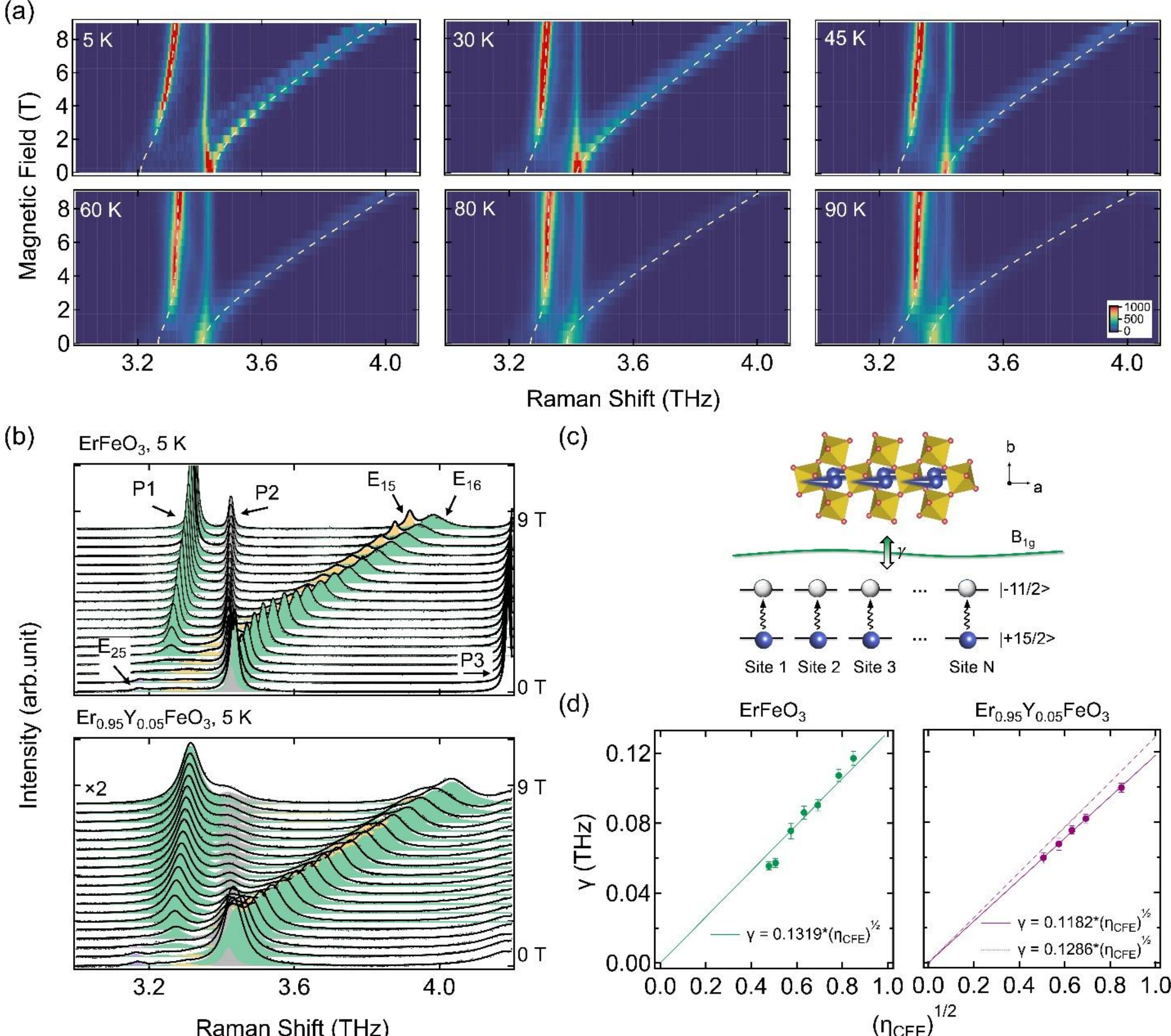

FIG. 4. Cooperative coupling between optical phonon and CFEs in $ErFeO_3$. (a) The avoided-crossing of the phonon-CFE coupling at different temperatures. The white dotted lines represent the fitting curves obtained by diagonalizing the 2×2 coupling matrix. (b) The magnetic-field dependent Raman spectra of undoped $ErFeO_3$(*top panel*) and Y doped $Er_{0.95}Y_{0.05}FeO_3$(*bottom panel*) at 5 K. (c) Schematic illustration of cooperative coupling among local JT ions mediated by virtual exchange of the optical phonon $P_1$. The atomic motions are derived from density functional theory calculations, primarily corresponding to the vibration of $Er^{3+}$ ions along the *a*-axis. (d) The linear scaling relation between coupling strength and the square root of effective ground-state population for the undoped $ErFeO_3$(*left panel*) and Y doped $Er_{0.95}Y_{0.05}FeO_3$(*right panel*). From left to right, the data points represent temperatures of: 90, 80, 60, 45, 30, 10, 5 K (*left panel*); and 80, 60, 45, 30, 5 K (*right panel*). The scaling relation is modeled using a linear function forced through the origin. Error bars represent the standard error derived from the 2 × 2 coupling matrix fitting.

Due to the strong SOC and low site symmetry in $ErFeO_3$, we do not observe a cooperative phase transition; instead, our results unravel the cooperative nature of optical phonon-CFEs coupling in the microscopic evolution connecting local JT interactions to cooperative structural instability, which is determined by the effective number of JT-active ions and the coherence of optical phonon field. The physical picture of such coupling is illustrated schematically in Fig. 4(c).

We now discuss the mechanism of $\sqrt{N}$-type scaling of the coupling strength between optical phonon and CFEs. This case belongs to two non-degenerate electronic states and a one-dimensional phonon, where the splitting of electronic states comes from both the orthorhombic crystal field effect and the Zeeman effect in magneto-Raman experiment. We therefore write the standard JT Hamiltonian as[1]:

$$H = \sum_{k} \hbar\omega(k)\left(a_k^{\dagger}a_k + \frac{1}{2}\right) + \sum_{k,n} \xi(k)e^{i\boldsymbol{k}\cdot\boldsymbol{R_n}} S_n^z\left(a_k + a_{-k}^{\dagger}\right) + \triangle \sum_{n} S_n^x \qquad (1)$$

where $a_k^{\dagger}$ and $a_k$ are the creation and annihilation operators of the phonon, $\omega(k)$ is the phonon frequency at wavevector $k$, $\xi(k)$ is the electron-vibrational coupling strength, $\boldsymbol{R_n}$ is the normal coordinate of the JT ion at site $n$, and $\Delta$ is the non-degenerate splitting mentioned above. The pseudospin operators $S_n^z$ and $S_n^x$ n take the form of Pauli matrices because the JT interaction term and the non-degenerate term do not commute with each other. Within the random phase approximation, the renormalized frequencies are found to be solutions of:

$$(\omega^2 - \omega_k^2)(\omega^2 - \omega_{CFE}^2) = \frac{4\omega^2\langle S_x\rangle\xi(k)^2}{\hbar^2} \qquad (2)$$

As a result, we can deduce the hybridization gap at the zero-detuning point $\omega_k = \omega_{CFE} = \omega_0$, which is $\Delta_{gap} = \frac{2}{\hbar}\sqrt{\frac{\Delta\langle S_x\rangle\xi(k)^2}{\hbar\omega_0}}$. The expected value of $\langle S_x\rangle$ follows tanh($\Delta/k_B T$), which is nothing but $\eta_{CFE}$. We thus achieve the observation of such linear scaling relation of the cooperative coupling between optical phonon and an ensemble of CFEs in (pseudo-)JTE, which also measures the electronic susceptibility corresponding to the two specific orbitals. Implicitly, Jahn-Teller models such as Eq. (1) assume a long coherent length across a macroscopic crystal region, which could break down in materials, this is why we usually consider uniform strain coupling. Strain destroys the Born-von Karman periodic boundary condition and develops out of the displacement of acoustic and optical phonons. Although acoustic phonons were thought to dominate, our findings highlight the contribution of optical phonons, which reconcile short-range order to complex structure that goes beyond simple ferro-distortive types.

The cooperative JTE can be mapped onto the Dicke model of quantum optics. If we drop the interactions apart from the Γ point in Eq. (1) (since the avoided-crossing suggests the leading term comes from $\boldsymbol{k} \approx 0$), then both the summation of k and phase factor are simplified. Consequently, the complex Hamiltonian boils down to an exactly solvable many-body problem: the Dicke model[37–39]. Within this framework, the optical phonon serves as a coherent bosonic field interacting with an ensemble of $N$ two-level systems (the $Er^{3+}$ CFEs). This mapping reveals that the cooperative JTE could be a solid-state analogue to the superradiant phase transition, with phonons playing the role of the cavity photons. Applying standard criteria from cavity quantum electrodynamics (cavity-QED), we find that at 5 K, the undoped system exhibits a cooperativity parameter $C = \frac{\gamma_{coll}^2}{\nu_{\mathrm{ph}}\nu_{CFE}} = 7.4$ and a normalized coupling strength of $\frac{2\gamma_{coll}}{\sqrt{\omega_{\mathrm{ph}}\omega_{CFE}}} = 10.7$ %, where $\nu$ denotes the peak full width at half maximum of phonon or CFE at zero field. These values confirm that the system resides firmly in the strong-coupling regime, where the interaction rate exceeds the individual decoherence rates.

This Dicke-model mapping suggests that certain cooperative JTE systems, specifically those characterized by strong optical phonon-CFEs coupling at the Γ point, may serve as solid-state quantum simulators for studying multi-mode Dicke physics[40–42]. Conversely, conceptual frameworks developed within cavity-QED may offer fresh insights into strongly correlated phenomena mediated by virtual phonon exchange. This is particularly relevant to the Kugel-Khomskii model, where vibronic interactions underpin superexchange mechanisms[43, 44]. The ability to tailor the (pseudo-)JTE strength thus offers a potential pathway to control macroscopic phases, such as Mott insulators or unconventional superconductors, a pursuit increasingly facilitated by advances in ultrafast spectroscopy. In the specific case of $ErFeO_3$, where driving selected phonons has been shown to transiently excite qAFM modes[45], our results suggest that the modulation of magnetic anisotropy via the $P_1$ phonon may offer a microscopic mechanism for coherent magnon excitation.

In summary, we have observed strong coupling between an optical phonon and CFEs in the pseudo JT system $ErFeO_3$, manifested as a prominent spectral avoided-crossing. By independently control of the temperature and Y-doping, we established a linear scaling relation between the coupling strength and the square root of the effective ground-state population. A 5 % substitution with the JT-inactive ion $Y^{3+}$ suppresses the hybridization gap far more strongly than expected from density scaling alone, highlighting the pivotal role of phonon coherence, which is often neglected in theoretical frameworks. We show that this $\sqrt{N}$-type scaling relation arises from cooperative coupling between a long-wavelength optical phonon field and an ensemble of CFEs within mean-field approximation, and we envision the observation of nonlinear scaling relation in JT systems with competitive order. Furthermore, when this strong coupling is dominated by the Γ point, the complex JT Hamiltonian can be approximately mapped onto the Dicke model in quantum optics, indicating the cooperative JTE may possess cooperative enhancement of symmetry breaking, which is similar to the $N^2$ enhancement of intensity in superradiant phase transition. The cooperative coupling competes with the impurity and defect effects, when local strains are reinforced synchronously, the entire structural instability occurs. We propose population control of vibronic cooperativity as a potent degree of freedom for the dynamical manipulation of macroscopic material properties.

## ACKNOWLEDGMENTS

We thank Xinwei Li for valuable discussions.

This work was supported by the National Key Research and Development Program of China (Grants No. 2020YFA0309200, 2021YFA1403800, 2024YFA1408700), the National Natural Science Foundation of China (Grants No. 12474475, 12374116, 12125404, T2495231, 12250008, 12188101), the Natural Science Foundation of Jiangsu Province (Grants No. BK20240057, BK20243011, and BK20233001), the Fundamental Research Funds for the Central Universities, the CAS project for Young Scientists in Basic Research (Grant No. YSBR-059), the fellowship from the China Postdoctoral Science Foundation (Grants No. 2024M751931, 2025M773408 and GZB20250786), the Jiangsu Funding Program for Excellent Postdoctoral Talent (Grants No.2025ZB376). M. B. is supported by the Research Foundation for Opto-Science and Technology, and the Japan Society for the Promotion of Science (JSPS) (Grant No. JPJSJRP20221202 and KAKENHI Grants No. JP24K21526, JP25K00012, JP25K01691, and JP25K01694).

## DATA AVAILABILITY

The data that support the findings of this article are openly available.